\documentclass[12pt]{iopart}

\ifx\pdfoutput\undefined
  \usepackage[dvips]{graphicx}
  \usepackage[dvips]{color}
\else
  \usepackage[pdftex]{graphicx}
  \usepackage[pdftex]{color}
\fi

\usepackage{iopams}
\usepackage{setstack}
\usepackage{latexsym}
\usepackage{units}
\usepackage{ulem}

\begin{document}

\title{Al/AlO$_{\rm {x}}$/Al-Josephson-junction-based microwave generators for weak-signal applications below 100 mK}

\author{Behdad Jalali-Jafari$^{1,2}$, Sergey V. Lotkhov$^1$ and Alexander B. Zorin$^1$}
\address{$^1$  Physikalisch-Technische Bundesanstalt, Bundesallee
100, 38116 Braunschweig, Germany}
\address{$^2$ Department of Microtechnology and Nanoscience (MC2), Chalmers University of Technology, SE-412 96, G\"oteborg, Sweden} \ead{Behdad.Jalalijafari@ptb.de}
\date{\today}

\begin{abstract}
An application-specific Josephson generator is being developed as an on-chip calibration source for single-microwave-photon detectors operating below $T \sim\unit[100]{mK}$ in the frequency range from $\unit[110]{}$ to $\unit[170]{GHz}$. The targeted calibration power corresponds to the photon detection rate of 0.01 to 10 counts per second. We analyze the Josephson oscillations within the framework of the tunnel junction microscopic (TJM) model and find the corrections to be made to a standard RSJ model of the junction, operating as a quasi-monochromatic source of microwaves. Experimentally, we implemented a number of aluminium-based two-junction interferometers connected to the leads via microstripline resistors made of partly oxidized titanium films. The $I$-$V$ curves  exhibit a behaviour ensuring generation of the microwave signal in the frequency range required. By setting the biasing point $\overline{V}$ and applying the magnetic field, one can tune the generator's frequency and power, respectively. A small oscillation linewidth, down to a few GHz, and a low level of black-body radiation emitted by hot parts of the circuit are predicted, using the experimental parameters.

\end{abstract}

\maketitle

\section{Introduction}

Dimensional scaling of superconductive circuit elements down to the nanometer range, accompanied by essentially weaker signals being processed, gives birth to a number of quantum devices operating below $T\sim\unit[100]{mK}$, manipulating single quanta of electrical quantities. For example, in electrical metrology, single-charge pumping \cite{PothierPump} provides a basic mechanism for the development of a capacitance standard \cite{KellerScience99,CamarotaMetrologia} and a current standard \cite{PekolaRevModPhys13}. In the superconducting qubits, single charge-, single flux- or single photon-states interfere to give rise to a variety of quantum algorithms \cite{Wendin2006,DevoretQubitReview04}.

The accurate operation of single-quantum circuits requires an exceptionally low background electromagnetic radiation which is quiescent in a wide frequency range on the utmost level of single ambient photons. For instance, the ultimate electron pumping accuracy was found to deteriorate over many orders of magnitude (the experimental uncertainty $\epsilon\sim10^{-8}$ greatly exceeding  the theoretical estimate $\epsilon \sim 10^{-20}$) due to a photon-assisted tunnelling phenomena \cite{KautzPAT00,PekolaEA10}. A possibility of significant improvement has been recently demonstrated for the  retention time of single electrons in a $\underline{{\rm S}}$ingle $\underline{{\rm E}}$lectron $\underline{{\rm T}}$unnelling (SET) trap using a high-quality
 mK-setup with double shielding of the sample space \cite{Kemppinen2011}. Furthermore, the radiation-induced generation of quasiparticles was considered to be one of the major energy relaxation mechanisms in the superconducting qubit systems (see \cite{Barends2011} and references therein).

Single-photon detection circuits operating in the microwave frequency range of up to a few hundred GHz and comprising an important part of a cryogenic toolbox at mK-temperatures have been a matter of study for many groups (see, e.g., \cite{Astafiev2002,Bozyigit2010,Chen2011,Oelsner2013}). Compared to single-photon detection in the optical and infrared ranges, the detection of microwave photons of relatively low energies requires components to be much more sensitive \cite{Astafiev2002,Bozyigit2010}. In our previous studies \cite{LotkhovNJP2011,LotkhovAPL2012}, we demonstrated the operation of a SINIS ($\underline{{\rm S}}$ = superconductor, $\underline{{\rm I}}$ = insulator, $\underline{{\rm N}}$ = normal metal) double-junction electron trap in its function as a microwave detector based on photon-assisted tunnelling. Featured by the energy scale $\sim\unit[0.1-1]{meV}$ inherent to the mesoscopic superconducting structures and weak back-action effects and being sensitive to single absorbed photons, this device shows great promise for applications, but requires calibration using a well-defined source of a weak microwave radiation.

Numerous implementations of microwave signal generators and local oscillators based on the Josephson junction arrays \cite{Sauvageau1987} or flux-flow circuits \cite{Mygind2002,Koshelets2013} were reported earlier. Successful experiments with weak-signal Josephson generators and on-chip devices under study were reported in \cite{VisscherThesis1996,Deblock2003,Billangeon2007}. For a wide range of applications, the single junction circuits, however, are considered (see, e.g., pp. 416-417 in \cite{LikharevBlueBook}) to produce a relatively weak signal, typically up to the level of nanowatts, and a relatively wide linewidth up to a few GHz. Moreover, a single junction component cannot be readily matched to an external circuit due to its low output impedance, typically on the level below $\unit[1]{\Omega}$, for the photolithographically-patterned junctions. Fortunately, these limitations should not be critical for the on-chip calibration of a microwave detector operating on the ultimate level of single photons and requiring a moderate emission power (and a moderate generation linewidth) only.

In the present work, we formulate the requirements, develop a model and report the experimental data of a tunable Josephson junction microwave source optimized for calibrating the SINIS detectors $prior$ $to$ $use$. Our analysis is focused on the sub-$\unit[100]{mK}$ operation of a two-junction interferometer co-fabricated with a compact network of on-chip  resistors, providing the controlled external impedance and designed to minimize the overheating of the circuit at low temperatures. Aluminium has been chosen for the superconductive material in order to match the frequency range and share the fabrication technique used for the SINIS detector \cite{LotkhovNJP2011}. By tuning a biasing point, i.e. average voltage $\overline{V}$, and adjusting a magnetic flux $\Phi_{\rm ext}$ through the interferometer loop, one can tune the generator's frequency ($f_{\rm J} = \overline{V}/\Phi_0,~\Phi_0 = h/2e$ is the flux quantum) and power ($\propto I_{\rm C}^2(\Phi_{\rm ext}),  I_{\rm C}$ is the critical current), respectively. The circuit parameters are evaluated with respect to a possibility of a narrow-band Josephson generation ($\delta f \ll f_{\rm J}$) with a sufficiently high amplitude (power).

\section{Requirements for a photon source for the single photon detector}

\begin{figure}[]
\centering%
\includegraphics[width=0.80\columnwidth]{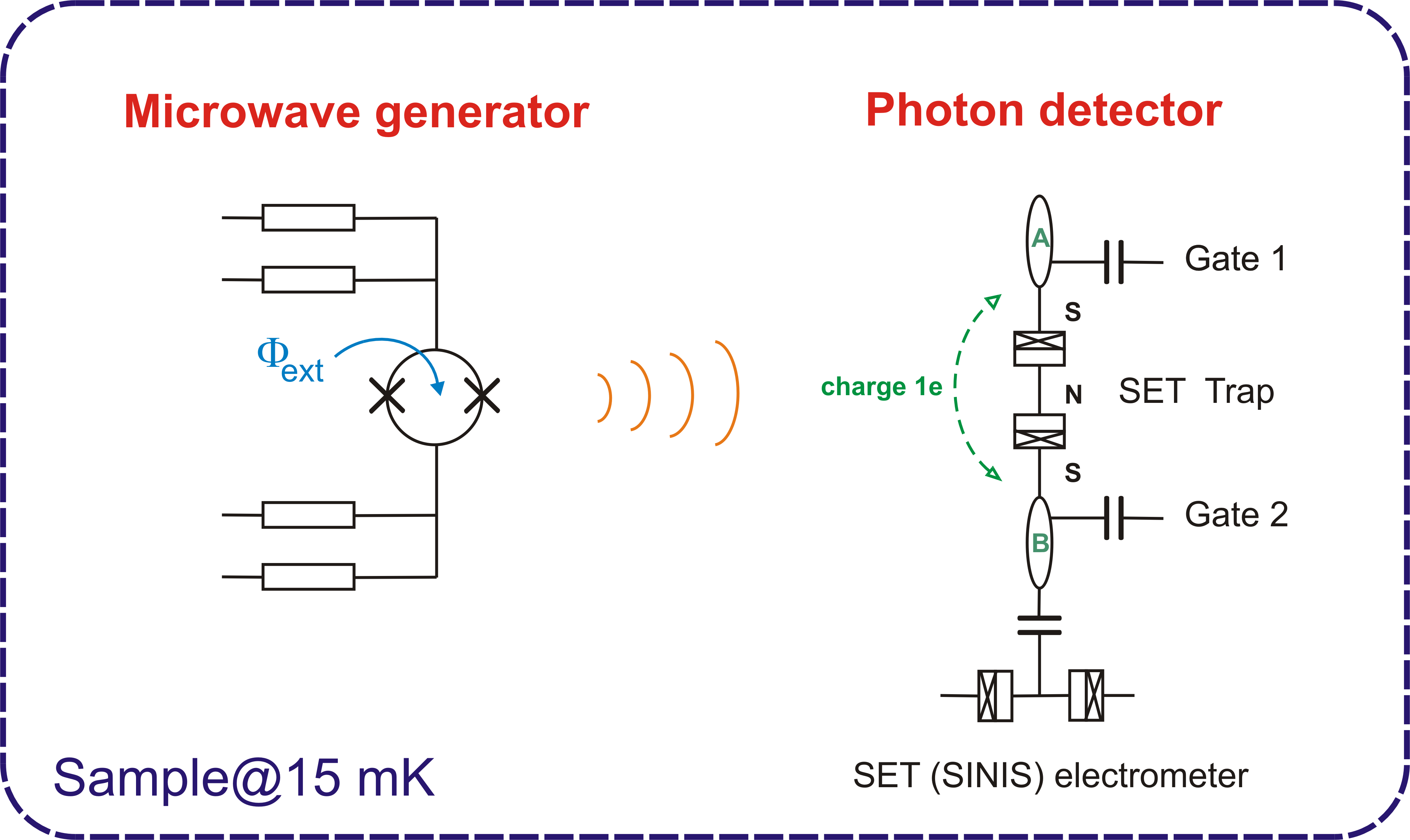}
\caption{(Color online) An equivalent circuit arrangement for the calibration experiment. The microwave source will be placed at some distance from the SINIS detector on the same chip. The detector consists of a SINIS-type SET trap and an SET electrometer used for counting the tunnelling events. Absorption of a photon leads to excitation and tunnelling of a quasiparticle across the SINIS double junction in either direction as shown by the dashed arrow. To enable such a process, the photon energy must exceed the activation threshold set by the superconducting gap energy $\Delta$ and the gate-tunable Coulomb energy of the tiny (typically, $\unit[0.03]{\mu m}\times\unit[0.1]{\mu m}\times\unit[1]{\mu m}$) N island.  }
\label{fig1GenDet}
\end{figure}

The goal of the present work is to develop a circuit solution and to experimentally realize a tunable Josephson junction generator for calibrating SET-based single photon detectors. The proposed arrangement for a complete experiment is shown in figure~\ref{fig1GenDet}.
Here we optimize the parameters of the Josephson generator to match the properties of the SINIS detector within a frequency range typical for the microwave background spectrum in a cryogenic setup at millikelvin temperatures. The details of operation of a SINIS detector can be found elsewhere \cite{LotkhovNJP2011,LotkhovAPL2012} and, in the current work, we restrict ourselves to considering the requirements they impose on the generators under construction.

For this purpose, we investigated the dark signal of a SINIS detector encapsulated in the sample holder within a dilution refrigerator of interest, both being the same as those used for characterizing the Josephson generators. In particular, we evaluated the photon counting rates for the background microwave radiation penetrating into the sample space from the warmer parts of the refrigerator due to a possible imperfection in the shielding. The detector sample was cooled down to $T \approx\unit[15]{mK}$ and installed into a single-wall-shielded, basically microwave-tight, but not vacuum-tight chip holder with all signal lines equipped at the mK-stage with a $\unit[1]{m}$-long Thermocoax$\texttrademark$ coaxial cable filter each \cite{ZorinThermocoax}.
The photon spectrum in the sample cavity is generally not well known and cannot be used for calibration of the detectors. On the other hand, it is a thorough characterization of the cryogenic environments, which, as mentioned above, motivates the development of the photon detection circuitry.

\begin{figure}[]
\centering%
\includegraphics[width=0.85\columnwidth]{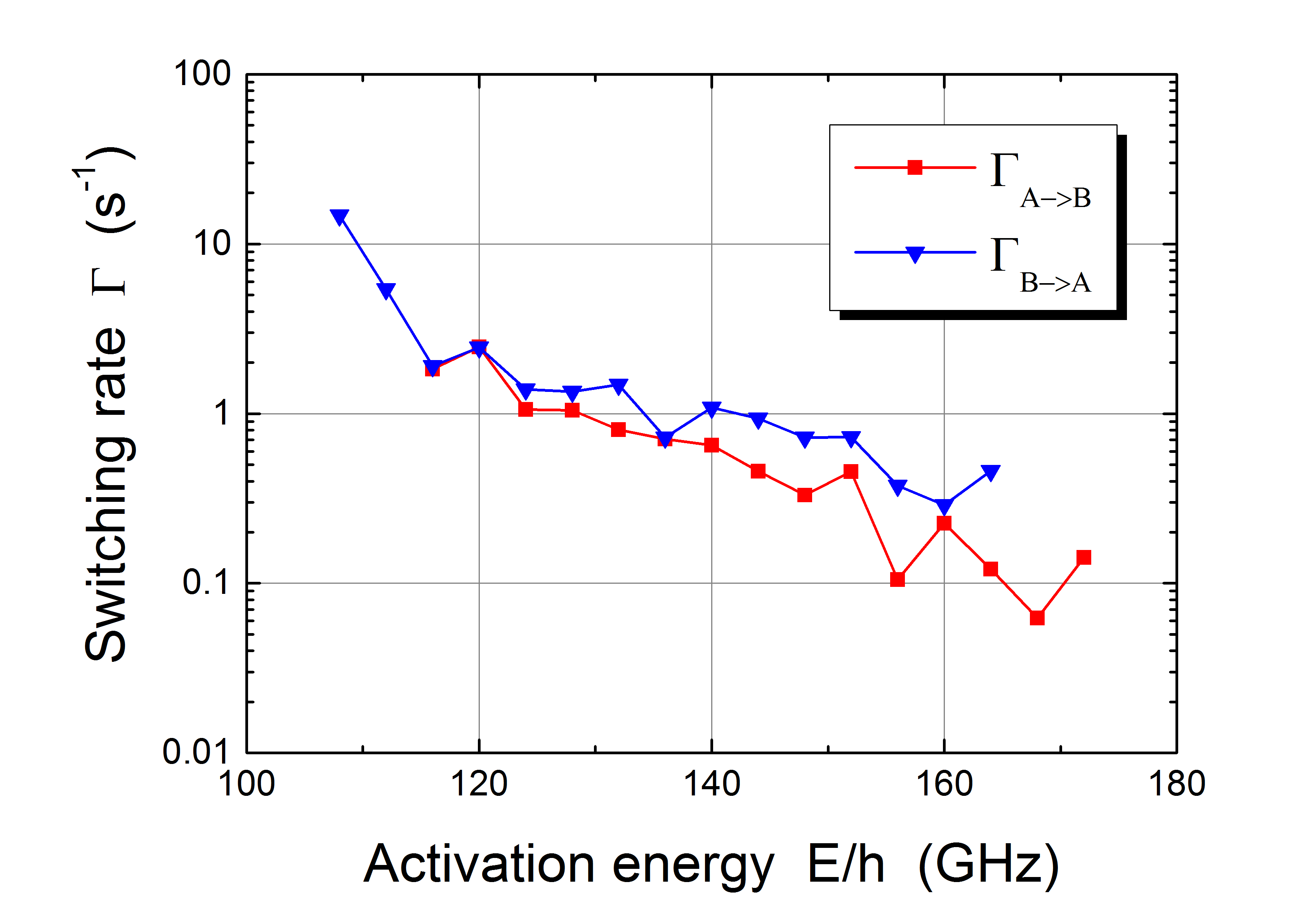}
\caption{(Color online) Charge-state switching rate vs. activation energy threshold value in a gate-tunable stand-alone SINIS detector. The rates are determined as inverse values of the average lifetime of an electron on the nodes A ($\Gamma_{{\rm A} \to {\rm B}} $) and B  ($\Gamma_{{\rm B} \to {\rm A}} $) shown in figure~\ref{fig1GenDet}. }
\label{fig2DetSwitchRate}
\end{figure}

The time resolution of measurements performed by our heavy-filtered, low-noise and high-resolution dc electronics is limited by the related $RC$ constants to about 100~Hz and the background charge drift reduces the full time available for the acquisition without re-tuning to few thousands of seconds. Therefore, the photon counting is assumed to receive meaningful statistics for the rates in the range of about 0.01 to about 10 counts per second. Figure~\ref{fig2DetSwitchRate} demonstrates, on a logarithmic scale, a threshold energy dependence of an electron switching rate between the nodes A and B in the SINIS detector shown on the right-hand side in figure~\ref{fig1GenDet}. The activation energy threshold was tuned using the gates~1 and 2, while the charge of the node B was monitored by the electrometer.

The data shown demonstrate the ability of the SINIS detector to discriminate the contribution of photons in the frequency range from about 110 to 170~GHz. The lower limit appears due to the background noise, whereas the upper limit is set by the Coulomb energy in the SINIS double junction. The absorption of higher-energy photons leads inevitably to the added counts and cannot be resolved in a spectral form. On the contrary, the detailed characterization in the lower-energy range, even if hampered by a too intensive background in our setup, could, in principle, be possible in a less-noisy system, as reported in \cite{Kemppinen2011}.

Within the frequency range of interest, we assume an intrinsically smooth spectrum curve and attribute the measured scatter to the properties of the SINIS detector (e.g., background charge drift induced instability of tuning, limited number of counts, etc.), which could, however, be improved in future experiments. For the present state-of-the-art technique, figure~\ref{fig2DetSwitchRate} also indicates that we need a rather moderate frequency resolution of our spectroscopy, viz.  on the level of about 5~GHz.

Finally, we can summarize the requirements for the calibration source to be able to produce a narrow-band microwave emission tunable in the frequency range $f$ = 110--$\unit[170]{GHz}$ with the linewidth $\delta f < \unit[5]{GHz}$ (prospectively,
$\delta f < \unit[1]{GHz}$) and of sufficient power to ensure, for very weak radiative coupling, about 0.01--10 photons per second to be absorbed by the detector.

\section{Theory and Models}

\subsection{Initial assumptions}

In a general representation (see, e.g., ch. 4 in \cite{LikharevBlueBook}), the Josephson oscillations of an autonomous junction show up as an infinite set of harmonics of a fundamental frequency $\omega_{\rm J} = 2\pi f_{\rm J}= (2\pi/\Phi_0)\overline{V}$, proportional to an average voltage across the junction $\overline{V}$.
Designing our microwave generator, we are aiming at the regime of small monochromatic Josephson oscillations, $V(t) \approx \overline{V} + V_{\rm 1} \sin(\omega_{\rm J}t + \chi)$, with the contribution of higher harmonics assumed to be negligibly small, i.e. the amplitudes $V_k \ll V_1 \ll \overline{V},$ $k = 2, 3, \dots$. The linewidth 2$\Gamma=2\pi\delta f$ of oscillations at frequency $\omega_{\rm J}$ determined by (presumably small) fluctuations of voltage \cite{LikharevBlueBook} should be sufficiently narrow, $\Gamma \ll \omega_{\rm J}$.  Under the given assumptions and according to the fundamental Josephson relation $\dot{\phi}=(2\pi/\Phi_0)V$,
the phase on the junction can be written as
\begin{equation}
\phi(t) = \omega_{\rm J}t - a \cos(\omega_{\rm J}t + \chi) + \phi_0, \label{eq:phase-time-dependence}
\end{equation}
where $a = (2\pi/\Phi_0) V_1/\omega_{\rm J} = V_1/\overline{V} \ll 1$ is expected to be a small parameter, $\phi_0$ is a constant phase. The amplitude of oscillations $V_1$ (and the value of $a$) depends on the working point $\overline{V}$ and the parameters of the circuit, and it can be assessed in different models of the Josephson junction. Finally, the experimental parameters must be optimized to provide a reasonable amplitude $V_1$, while keeping the requirement  $V_1 \ll \overline{V}$.

\subsection{Oscillation amplitude: RSJ Model}

The most simple $\underline{{\rm R}}$esistively $\underline{{\rm S}}$hunted $\underline{{\rm J}}$unction (RSJ) model \cite{McCumber,Stewart} (see also \cite{LikharevBlueBook}) is associated with the circuit equation
\begin{equation}
 C \left(\frac{\Phi_0}{2\pi}\right) \ddot{\phi} + \frac{1}{R_{\rm S}} \left(\frac{\Phi_0}{2\pi}\right) \dot{\phi} + I_C \sin \phi= I_{\rm bias} \label{eq:RSJ model eq}
\end{equation}
 which includes the capacitive current (proportional to the junction capacitance $C$), the normal current through the shunt resistance $R_{\rm S}$ and the Josephson supercurrent $I_{\rm C} \sin\phi$. The oscillation amplitude $V_1$ can be easily found by substituting (\ref{eq:phase-time-dependence}) with small $a$ into the equation of motion (\ref{eq:RSJ model eq}) and then excluding variable $\chi$, thus achieving
\begin{equation}
V_1 = a \overline{V} = \frac{V_{\rm C}}{\sqrt{1 + (\beta_{\rm C} \frac{\overline{V}}{V_{\rm C}})^2}}, \label{eq:amplitude}
\end{equation}
where $\beta_{\rm C} \equiv \frac{2\pi}{\Phi_0}I_{\rm C}R_{\rm S}^2C$ is the McCumber-Stewart damping parameter and $V_{\rm C} \equiv I_{\rm C}R_{\rm S}$ the value of the characteristic voltage. The amplitude of the second harmonic can be evaluated, in this case, as
\begin{equation}
V_2 \approx V_1 \frac{V_{\rm C}/2\overline{V}}{\sqrt{1 + (\beta_{\rm C} \frac{2\overline{V}}{V_{\rm C}})^2}} \ll V_1. \label{eq:amplitudeV2}
\end{equation}
Since $V_2 \sim a^2 \overline{V}$, the second harmonic can be neglected for sufficiently small $a$. Similar relations can be obtained also for all higher harmonics $k = 3, 4, \dots$, i.e. $V_k \sim a^k \overline{V}$.

In the RSJ model, the inherent properties of tunnel junctions, i.e. strongly nonlinear quasiparticle current $I_{\rm qp}(V)$ and frequency dispersion of the supercurrent, are omitted. These can be adequately taken into account within the framework of the $\underline{{\rm T}}$unnel $\underline{{\rm J}}$unction $\underline{{\rm M}}$icroscopic (TJM) model developed by Werthamer \cite{Werthamer} and Larkin and Ovchinnikov \cite{LarkinOvchinnikov}.

\subsection{Corrections by TJM Model}

Within the framework of the TJM model (see general microscopic formulas for tunnelling current \,(2.2) and (2.3) in \cite{LikharevBlueBook}) the circuit equation for an autonomous Josephson junction can be presented in the form (\ref{eq:RSJ model eq}) with the modified supercurrent term
\begin{equation}
I_C \sin \phi \quad \rightarrow \quad {\rm Im}\,\sum_{m,n}A_m A_n I_p[(n+1)\omega_{\rm J}] e^{i[(m+n+1) \omega_{\rm J}t + \phi_0]}
 \label{eq:TJM model eq Ip}
 \end{equation}
and an additional term describing quasiparticle current,
\begin{equation}
I_{\rm quasiparticle} = {\rm Im}\,\sum_{m,n}A_m A_n^* I_q[(n+1)\omega_{\rm J}] e^{i(m-n) \omega_{\rm J}t}.
 \label{eq:TJM model eq Iq}
 \end{equation}
Here $A_n$, $n=0, \pm1, \pm2, \dots$, are the complex Fourier coefficients in the expansion
\begin{equation}
e^{i\phi(t)/2} = \sum_{n} A_n e^{i[(n+0.5) \omega_{\rm J}t + 0.5 \phi_0]},
 \label{eq:TJM model An}
 \end{equation}
whereas $I_{\rm p}(\omega)$ and $I_{\rm q}(\omega)$ are the complex functions of frequency reflecting the dispersion of the Cooper-pair current and the quasiparticle current, respectively (corresponding plots of $I_{\rm p,q}(\omega)$ are given in, e.g., figure \,2.2 of \cite{LikharevBlueBook}). The shapes of these functions depend on the temperature
and properties of the junction electrodes (in particular their superconducting energy gap $\Delta$) and
the barrier transparency. Some of their components have a clear physical interpretation. For example, the term  ${\rm Im}I_{\rm q}(\omega)|_{\omega = 2e\overline{V}/\hbar} \equiv I_{\rm qp}(\overline{V})$ coincides with the bare quasiparticle $I$-$V$ curve of the tunnel junction, whereas the value ${\rm Re}I_{\rm p}(0) = I_{\rm C}$, i.e., represents the critical current. The components ${\rm Re}I_{\rm q}$ and ${\rm Im}I_{\rm p}$ are the effective reactance of the tunnel junction associated with quasiparticle current and the so-called dissipative quasiparticle-Cooper-pair interference term, respectively \cite{Poulsen}.

For the phase dependence given by (\ref{eq:phase-time-dependence}) coefficients $A_n$ take the values
\begin{equation}
A_0 = e^{i\phi_0/2},\qquad A_{\pm 1} = i \frac{a}{4} e^{\pm i(\chi + \phi_0/2)},\qquad A_{n}=0,\quad |n|>1.
 \label{eq:values An}
 \end{equation}
Then the quasiparticle and Cooper-pair components of tunnelling current can be presented explicitly as a sum of dc current and small ($\propto a$) ac current oscillating with frequency $\omega_{\rm J}$. The balance of the latter current with a Maxwell's displacement current through the junction capacitance and ac current through a parallel resistive shunt yields the amplitude of Josephson voltage oscillations on the junction
\begin{equation}
V_1 = \frac{\Phi_0}{2 \pi} \omega_{\rm J}
\frac{\left[ {\rm Re}^2 I_{\rm p}(\omega_{\rm J}/2)+ {\rm Im}^2 I_{\rm p}(\omega_{\rm J}/2)\right]^{1/2}}
{(Q^2 + S^2)^{1/2}}, \label{eq:amplitude TJM}
\end{equation}
where
\begin{equation}
Q = {\rm Im} \left[I_{\rm q}(\omega_{\rm J}/2)+ I_{\rm q}(3\omega_{\rm J}/2)\right]/4 + (\Phi_0/2 \pi)\omega_{\rm J}/R_{\rm S}, \label{eq:amplitude TJM2}
\end{equation}
\begin{equation}
S= {\rm Re}  \left[I_{\rm q}(\omega_{\rm J}/2)- I_{\rm q}(3\omega_{\rm J}/2)\right]/4 - (\Phi_0/2 \pi)\omega_{\rm J}^2C. \label{eq:amplitude TJM3}
\end{equation}
We note that in a special case of an ohmic quasiparticle current (for example, when the temperature is close to the critical one, $T \lesssim T_c$), $I_{\rm q}(\omega) = i(\Phi_0/\pi)\omega/R_{\rm N}$ and the Cooper-pair current taken in a simple form $I_C \sin\phi$ (resulted from $I_{\rm p}(\omega) = {\rm Re}I_{\rm p}(0) = I_{\rm {\rm C}}$), the formula (\ref{eq:amplitude TJM}) reduces to the result (\ref{eq:amplitude}) obtained from the RSJ model.

An RSJ model relation, however, can be derived also for the most interesting case of sufficiently low temperature $T \ll T_c$, and dc voltage $\overline{V}$ falling into the range from about $2\Delta/3e$ up to $2\Delta/e$ (i.e., the Josephson frequency $\omega_{\rm J}/2\pi$ ranging from about $65\,$GHz up to $200\,$GHz for Al electrodes). The functions $I_{\rm q}(\omega)$ and $I_{\rm p}(\omega)$ can be roughly estimated (see the plots of $I_{\rm p,q}(\omega)$ in figure \,2.2 of \cite{LikharevBlueBook}), so the terms entering the equations (\ref{eq:amplitude TJM}-\ref{eq:amplitude TJM3}) are simplified as

\begin{equation}
\left[ {\rm Re}^2 I_{\rm p}(\omega_{\rm J}/2)+ {\rm Im}^2 I_{\rm p}(\omega_{\rm J}/2)\right]^{1/2} \approx {\rm Re} I_{\rm p}(\omega_{\rm J}/2) \approx I_{\rm C},
\label{eq:simlified amplitude TJM 1}
\end{equation}
\begin{equation}
Q \approx {\rm Im} I_{\rm q}(3\omega_{\rm J}/2)/4 + (\Phi_0/2 \pi)\omega_{\rm J}/R_{\rm S} = (\Phi_0/2 \pi)\omega_{\rm J}(\tilde{R}_{\rm qp}^{-1}+R_{\rm S}^{-1}),
\label{eq:simlified amplitude TJM 2}
\end{equation}
\begin{equation}
S \approx -I_{\rm C}/4 - (\Phi_0/2 \pi)\omega_{\rm J}^2C = -(\Phi_0/2 \pi)\omega_{\rm J}^2(\tilde{C}_{\rm qp} + C). \label{eq:simlified amplitude TJM 3}
\end{equation}
The corrections to the shunt conductance (\ref{eq:simlified amplitude TJM 2}) and the junction capacitance (\ref{eq:simlified amplitude TJM 3}) due to quasiparticle current are
\begin{equation}
\tilde{R}_{\rm qp}^{-1} \approx (3/4)R_{\rm N}^{-1} {\quad \rm and \quad} \tilde{C}_{\rm qp} \approx I_{\rm C} (\pi/2\Phi_0)\omega_{\rm J}^{-2},
\label{eq:correction R-C}
\end{equation}
respectively. Combining (\ref{eq:amplitude TJM}-\ref{eq:amplitude TJM3}) with (\ref{eq:simlified amplitude TJM 1}-\ref{eq:correction R-C}), one can finally see that for evaluating the amplitude of Josephson oscillations, one can still apply (\ref{eq:amplitude}) with effective parameters
\begin{equation}
V_{\rm C}^* \equiv I_{\rm C}R^*  {\quad \rm and \quad} \beta^*_{\rm C}(\omega_{\rm J}) \equiv \frac{2\pi}{\Phi_0}I_{\rm C}R^{*2}C^*(\omega_{\rm J}),
\label{eq:correction Vc beta}
\end{equation}
where the effective values are
\begin{equation}
R^*=(R_{\rm S}^{-1} + \tilde{R}_{\rm qp}^{-1})^{-1}  {\quad \rm and \quad} C^*(\omega_{\rm J})=C+\tilde{C}_{\rm qp}(\omega_{\rm J})\label{eq:effective R C}.
\end{equation}

\subsection{Biasing circuit layout and peculiarities of $I$-$V$ curve}

\begin{figure}[]
\centering%
\includegraphics[width=0.80\columnwidth]{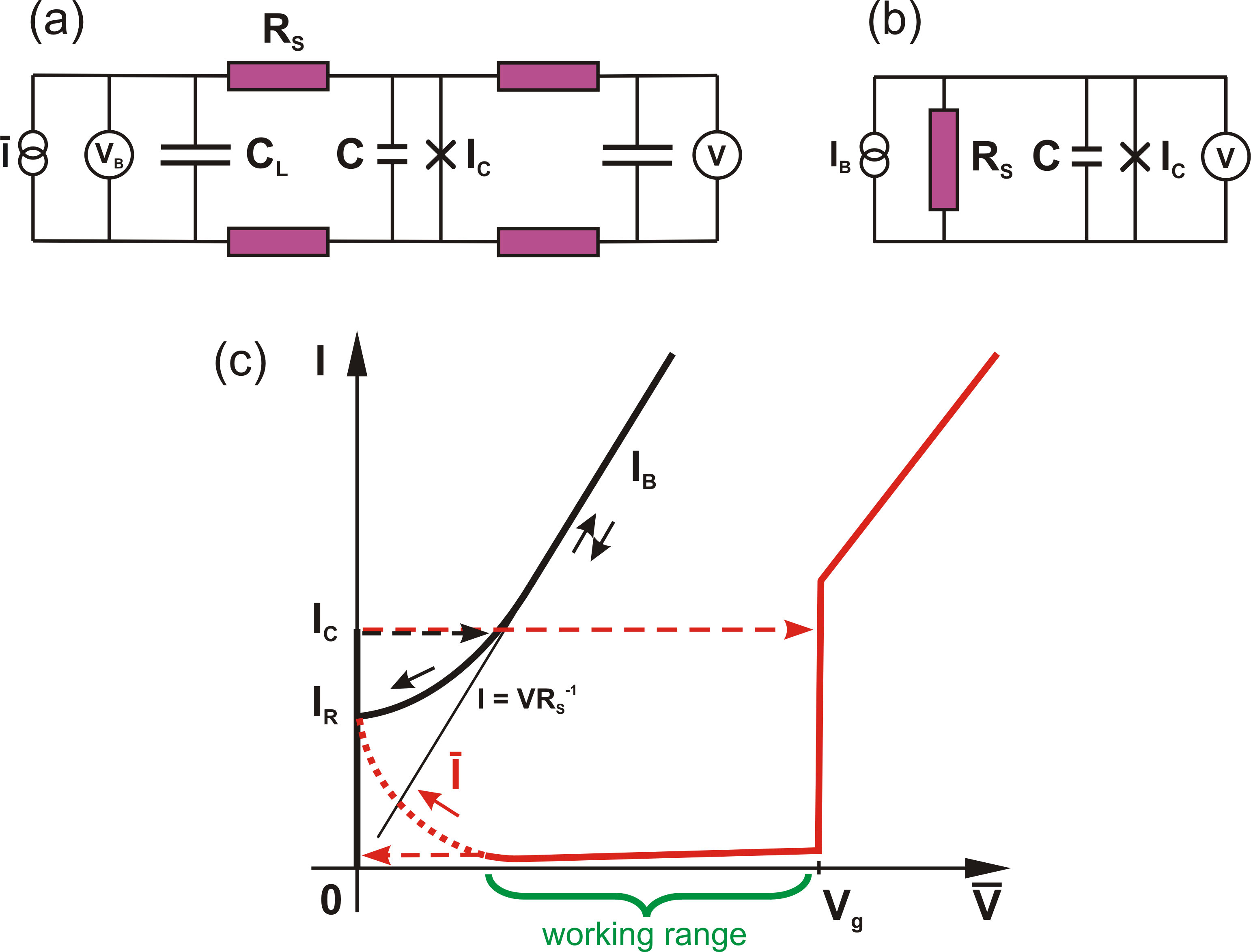}
\caption{(Color online) (a) An experimental circuit based on an unshunted Josephson junction. The capacitance $C_{\rm L}$ accounts for a relatively large lead capacitance which provides a dc voltage bias for the circuit (the Thevenin configuration). (b) The equivalent transformation to the Norton configuration. (c) A general view of the hysteretic $I$-$V$ curves of a shunted, $I_{\rm B}$-$\overline{V}$,  and unshunted, $\overline{I}$-$\overline{V}$, tunnel junction. The dashed arrows indicate the directions of  switching and retrapping processes, whereas the dotted segment is not accessible in measurements.}
\label{fig3ExpCircuit}
\end{figure}

The dynamics of a Josephson junction strongly depends on the external impedance \cite{LikharevBlueBook}. In experiments with small Al junctions \cite{VisscherThesis1996,Deblock2003,Billangeon2007,Steinbach2001}, a significant contribution is to be avoided which arrives from the on-chip biasing leads with a generally not well-defined impedance. Furthermore, the amplitude of oscillations vanishes with increasing values of the capacitance $C$ and the damping parameter $\beta_{\rm C}$, see (\ref{eq:amplitude}). In order to minimize the capacitive contribution from the leads, $C_{\rm L} \sim \unit[0.1]{pF}$, we followed the approach developed in \cite{Deblock2003,Billangeon2007} and biased the junction via a compact four-terminal network of reasonably high-ohmic resistors, $R_{\rm S} \sim \unit[1]{k\Omega} \gg (\omega_{\rm J}C_{\rm L})^{-1} \sim \unit[10]{\Omega}$, as shown in figure~\ref{fig3ExpCircuit}(a). The current feed $\overline{I}$ was provided through one pair of the resistors, while the voltage $\overline{V}$ was measured across the other one.

For the purpose of ac modelling, the Thevenin-type, i.e. voltage-bias, configuration shown in figure~\ref{fig3ExpCircuit}(a) with $V_{\rm B} = \overline{V} + 2R_{\rm S}\overline{I}$ can be transformed to the standard Norton-type circuit, see figure~\ref{fig3ExpCircuit}(b), with the current bias $I_{\rm B} = \overline{I} + \overline{V}/R_{\rm S}$ including a virtual (for the purpose of modelling only) dc current term through the shunt. Accordingly, as shown in figure~\ref{fig3ExpCircuit}(c), the experimental dc current $\overline{I}$ can be interpreted as a difference between the model current $I_{\rm B}$ and the fictive dc current term, $\overline{V}/R_{\rm S}$. For voltages $\overline{V} \ge V_{\rm g}$, the gradually vanishing dc part of the supercurrent is superimposed on a nonlinear quasiparticle-current branch.

From figure~\ref{fig3ExpCircuit}(c), it is obvious that the practical upper range of operation frequency $f_{\rm J}$ is basically set by the gap frequency $f_{\rm g} \equiv V_{\rm g}/\Phi_{\rm 0} = 4\Delta/h \approx 200$\,GHz for Al junctions. At a larger bias ($\overline{V} > V_{\rm g} = 2\Delta/e$) the strong dc quasiparticle current leads to the intensive noise and overheating of the junction. The lower frequency limitation is defined by the requirement of smallness of higher harmonics  (\ref{eq:amplitudeV2}), the assumptions made in section~ 3.1. This condition is safely fulfilled outside the area of the supercurrent cusp (with a width $\sim V_C$ \cite{Steinbach2001}), where $\overline{I_{\rm S}} \to 0$ and the $I$-$V$ curve almost restores the zero or even positive slope (see figure~\ref{fig3ExpCircuit}(c)). The condition for low $\overline{V}$ is less restrictive for the circuits with high  $\beta_{\rm C} >1$, see (\ref{eq:amplitude}).

Another limitation, which can become even more restrictive in experiments with circuits having a bias as shown in figure~\ref{fig3ExpCircuit}(a), is set by the instability of the $\overline{I}$-$\overline{V}$ curve at low currents and voltages with respect to retrapping the Josephson phase $\phi(t) \to \phi_0 = \arcsin(\overline{I}/I_{\rm C})$, i.e. switching of the junction to its zero-voltage state, as shown in figure \ref{fig3ExpCircuit}(c). The value of retrapping current $I_{\rm R}$ generally depends on the frequency-dependent damping in the circuit which is basically not well characterized in a wide frequency range, due to the contribution of quasiparticle current and signal leads. Therefore, it appears to be most practical to withdraw the points below the retrapping threshold directly from the experimental data.

\subsection{Thermal fluctuations and shot noise}

An important issue of operating a Josephson generator below $T \sim\unit[100]{mK}$ appears to be a weak electron-phonon coupling in metallic films in this temperature range \cite{Wellstood89}. A dramatic increase in the local electron temperature $T_{\rm e}$ occurring due to appreciable power dissipation in small volumes results in enhanced black-body radiation from hot parts of the circuit, $j=\sigma T_{\rm e}^4$, where $j$ is the radiation power per unit area and $\sigma \approx \unit[5.7 \times 10^{-8}]{Wm^{-2}K^{-4}}$ the Stefan-Boltzmann constant, as well as an increase of voltage fluctuations leading to widening of the Josephson oscillation linewidth \cite{LikharevBlueBook} which we address below.

In microscale bodies like compact shunting resistors or small metal islands, the electron temperature  might considerably exceed  the phonon (lattice) temperature $T_{\rm ph} \sim T$, as indicated by the one-fifth power law \cite{Wellstood94}:

\begin{equation}
T_{\rm e} =\left[T_{\rm ph}^5+\frac{P}{\Sigma\Omega}\right]^{1/5}. \label{eq:ElecTemp}
\end{equation}
Here, $P$ is the Joule power dissipated in the given volume $ \Omega$  and $\Sigma$ is a material-dependent electron-phonon coupling constant. For example, considerable electron overheating $T_{\rm e} \sim \unit[1]{K}$ should be expected, if the model circuit shown in figure \ref{fig3ExpCircuit}(b) is realized straightforwardly, allowing the dc current dissipation in the parallel resistor $R_{\rm S}$, $P=\overline{V}^2/2R_{\rm S} \sim \unit[5]{nW}$, for the parameters described in the experimental section of this article. On the other hand, a negligibly small temperature increase is expected if the on-chip circuitry includes a serial RC-shunt in parallel to the Josephson junction (instead of a parallel shunting resistor) and it did not conduct at dc (see, e.g., \cite{Steinbach2001}). In the case of our circuit with low dc current through the bias resistors (figure~\ref{fig3ExpCircuit}(a)), the dissipated power is also low.

For estimating the linewidth of oscillations we have to consider two different sources of noise contribution at low frequencies: the first one is the thin film resistor  (generating Johnson-Nyquist noise) and the second one is the tunnel junction. The noise originating from the tunnel junction is dominated by shot noise in the regime where $e\overline{V}\gg k_{\rm B}T$. Therefore the total halfwidth $\Gamma _{\rm T}$ can be expressed as
\begin{equation}
\Gamma _{\rm T} =\Gamma _{\rm thermal}+\Gamma _{\rm shot},\label{eq:totalGamma}
\end{equation}
the former halfwidth for the bias $V_C\ll\overline{V}<V_g$ can be written as \cite{LikharevBlueBook}:
\begin{equation}
\Gamma _{\rm thermal} =\pi\left(\frac{2e}{\hbar}\right)^2\frac{R_{\rm d}^2}{R_{\rm S}}k_{\rm B}T_{\rm e}\approx\pi\left(\frac{2e}{\hbar}\right)^2R_{\rm S}k_{\rm B}T_{\rm e},\label{eq:linewidth1}
\end{equation}
where $R_{\rm d}\approx R_{\rm S}$  is the effective differential resistance (in equivalent configuration figure~\ref{fig3ExpCircuit}(b)) and $k_{\rm B}$ is the Boltzmann constant.
The latter halfwidth can also be estimated as \cite{LikharevBlueBook}:
\begin{equation}
\Gamma _{\rm shot} \approx \pi R_{\rm S}^2\left(\frac{2e}{\hbar}\right)^2 S_{\rm I}(0),\label{eq:linewidth2}
\end{equation}
where $ S_{\rm I}(0)$ is the current spectral density which for $e\overline{V}\gg k_{\rm B}T$ can be approximated as
\begin{equation}
S_{\rm I}(0)\approx \left(\frac{e}{2\pi}\right) {\rm Im} I_{\rm q}\left(\frac{\omega_{J}}{2}\right) {\quad \rm with \quad} {\rm Im} I_{\rm q}\left(\frac{\omega_{J}}{2}\right) \approx I_{\rm qp}\left(\overline{V}\right), \label{eq:CurrentSpectralDensity}
\end{equation}
where $I_{\rm qp}\left(\overline{V}\right)$ represents the residual quasiparticle current, typically suppressed below the gap by a few orders of magnitude.

\section{Experiment}

\subsection{Layout and fabrication}

\begin{figure}[]
\centering%
\includegraphics[width=0.80\columnwidth]{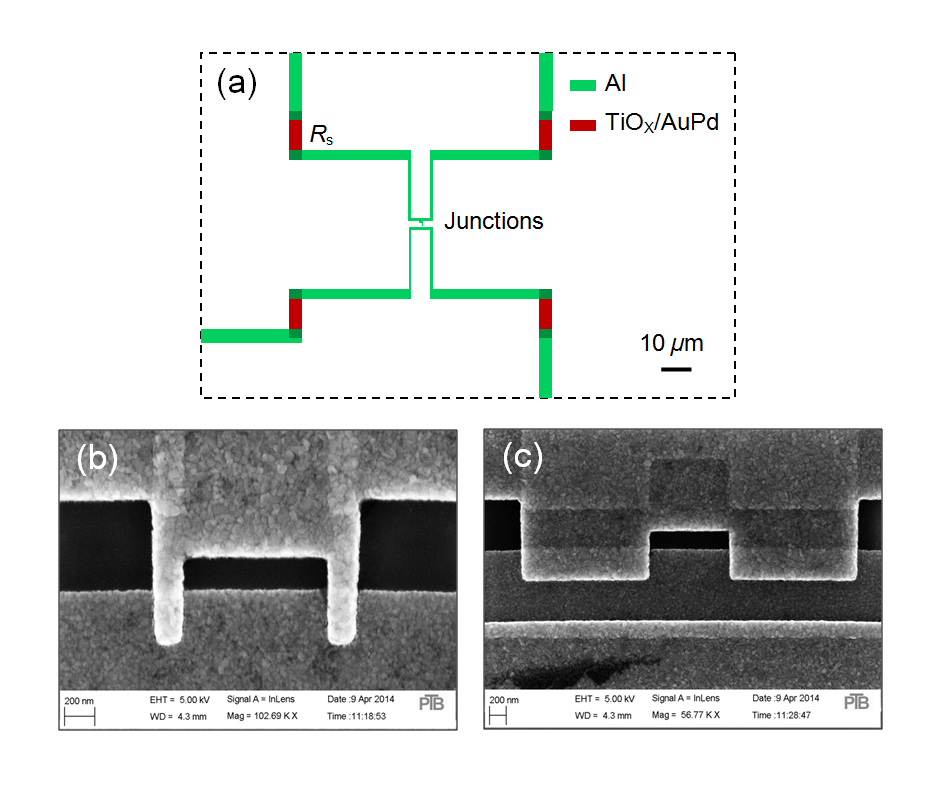}
\caption{(Color online) (a) Layout of the experimental devices consisting of two evaporation masks: (1) a set of four thin film resistors $R_{\rm S}$ (shown in red) and (2) a small-junction interferometer (in the centre) connected to the leads (green). (b) and (c) SEM images of type A and B devices, respectively. See the text for more details.}
\label{fig4layout-SEM}
\end{figure}

The basic 4-point terminal layout of our circuit is shown in figure~\ref{fig4layout-SEM}(a) and it was common for all 7 different devices studied. The dimensions of elements were designed to enable a comparison in a wide range of circuit parameters. All devices were fabricated on the same thermally oxidized Si substrate using a two-step e-beam lithography process. The first e-beam lithography was performed for the thin film resistors deposited through a PMMA/Ge/MMA(8.5)MAA mask with 0.1/0.03/0.6 $\unit{\mu m}$ thicknesses, respectively. The resistors with two different dimensions of $\unit[5]{\mu m} \times\unit[10]{\mu m}$ and $\unit[5]{\mu m} \times\unit[30]{\mu m}$ were fabricated by evaporating a $\unit[30]{nm}$ Ti film at a low oxygen pressure, $P_{\rm O2}\approx\unit[1 \times 10^{\rm{-6}}]{mbar}$, in the evaporation chamber. The oxidation during deposition was used to form the Ti film of higher resistivity \cite{LotkhovUltraHigh}, which was necessary to approach the target values for $R_{\rm S} \le \unit[1]{k\Omega}$, and which also prevented superconductivity of Ti at low temperature. The evaporation of the Ti film was followed $in$-$ situ$ by deposition of a $\unit[5]{nm}$-thin AuPd film, in order to ensure a low contact resistance to the leads of Al evaporated in the next steps.  At the final stage of fabrication, the film of AuPd was removed by Ar sputtering in some chips, using Al as an etching mask, in order to approach the resistivity of Ti film. The resistance measurements at $\unit[15]{mK}$ showed average values of $R_{\rm S}\approx\unit[180]{\Omega}$ $(\unit[330]{\Omega})$ for the shorter and $R_{\rm S}\approx\unit[550]{\Omega}$ $(\unit[1000]{\Omega})$ for the longer resistive pieces before (after) the removal of AuPd, respectively. These resistance values are well below the resistance quantum $R_{\rm Q}= h/4e^2\approx\unit[6.45]{k\Omega}$, which helps to avoid quantum fluctuations of the Josephson phase \cite{KochQuanNoise} and, therefore, widening of the linewidth of oscillations.

The second e-beam lithography was used for defining the Josephson junctions and the connection leads. Al junctions were fabricated using the shadow evaporation technique \cite{Dolan} and a similar PMMA/Ge/MMA(8.5)MAA mask of 0.1/0.05/1.2 $\unit{\mu m}$ thicknesses, respectively. The structures were designed in a two-junction-interferometer form which enables tuning the critical current and, therefore, the radiation power by applying a magnetic field. The first and the second Al layers ($\unit[18]{nm}$/$\unit[60]{nm}$) were evaporated to the substrate tilted to the angles $-17^\circ$/$+17^\circ$. The first layer was oxidized at $\unit[25]{Pa}$ for $\unit[10]{min}$ to produce a tunnel barrier, giving rise to a reasonably high critical current density $J_{\rm C}\approx\unit[2.5]{\mu A/\mu m^2}$, while keeping low the leakage currents $I_{\rm qp}\left(\overline{V}\right)$ at $T\ll T_C$ which are up to two orders of magnitude below $I_{\rm C}$. Two greatly different junction sizes were implemented (see figures~\ref{fig4layout-SEM}(b) and (c)): small junctions (type A) with the total area of two junctions $A\approx\unit[0.1]{\mu m^{2}}$, the estimated total capacitance of $C\approx\unit[5]{fF}$ and the total normal resistance $R_{\rm N}\approx\unit[900]{\Omega}$; and large junctions (type B) with  $A\approx\unit[1]{\mu m^{2}}$, $C\approx\unit[50]{fF}$ and $R_{\rm N}\approx\unit[100]{\Omega}$. The inner loop size of $\unit[1]{\mu m}\times \unit [0.28]{\mu m}$ was the same for both types, corresponding to one flux quantum $\Phi_0$ for the external magnetic field $B \approx \unit[7]{mT}$.

\subsection{DC characterization}

\begin{figure}[]
\centering%
\includegraphics[width=0.9\columnwidth]{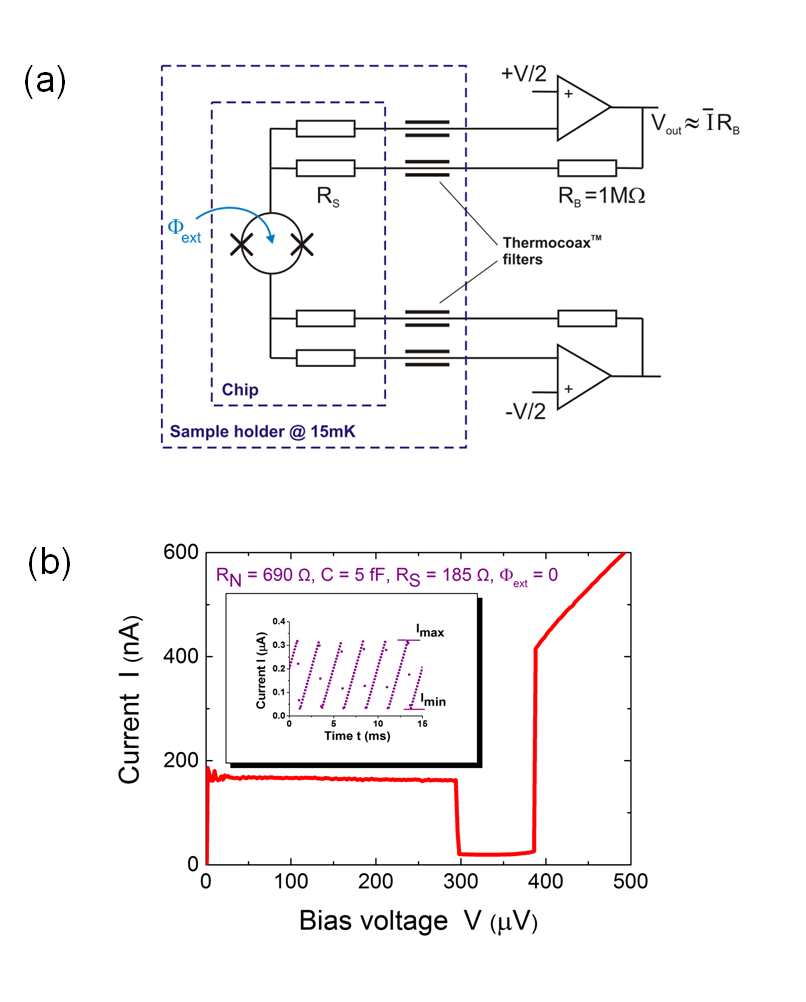}
\caption{(Color online) (a) An equivalent measurement circuit based on a transimpedance amplifier as an active (feedback-controlled) dc voltage source and a current meter. (b) Example of a measured $I$-$V$ curve for a type A device. For voltages in the range $\unit[300]{\mu V} < V < \unit[380]{\mu V}$ the biasing points are dc-fixed and can be used for the microwave generation. The biasing points below $V \approx \unit[300]{\mu V}$ are unstable and the current measurement $I$ produces the average of the switching current $I_{\rm max} \approx I_{\rm C}$ and the retrapping current $I_{\rm min }$ values shown in the inset. Inset shows a low-frequency autonomous cycling of the current through the junction in the unstable range.}
\label{fig5MeasIVcurve}
\end{figure}

The $I$-$V$ curves of the Josephson-junction circuits were measured in the dilution refrigerator, in the setup described above, using room temperature electronics built around a transimpedance amplifier drawn in figure~\ref{fig5MeasIVcurve}(a). Figure~\ref{fig5MeasIVcurve}(b)  shows a typical $I$-$V$ curve measured for a type A device at $T\approx\unit[15]{mK}$. Due to the feedback circuitry, the dc voltage setting refers to the junctions directly and independently on the line and the shunt resistance. Therefore, within the useful part of the branch with the stable voltage and current, see also figure~\ref{fig3ExpCircuit}(c), a direct frequency tuning is facilitated by setting $\overline{V}$ ($= f\times \Phi_0$) directly. At lower voltages, the process of retrapping to the zero-voltage state, followed by the counter action of the feedback circuitry, leads to a self-maintained low-frequency cycling of the dc current across the junction as shown in inset figure~\ref{fig5MeasIVcurve}(b). No stable generation is possible in this voltage (and frequency) range.

\begin{figure}[]
\centering%
\includegraphics[width=0.9\columnwidth]{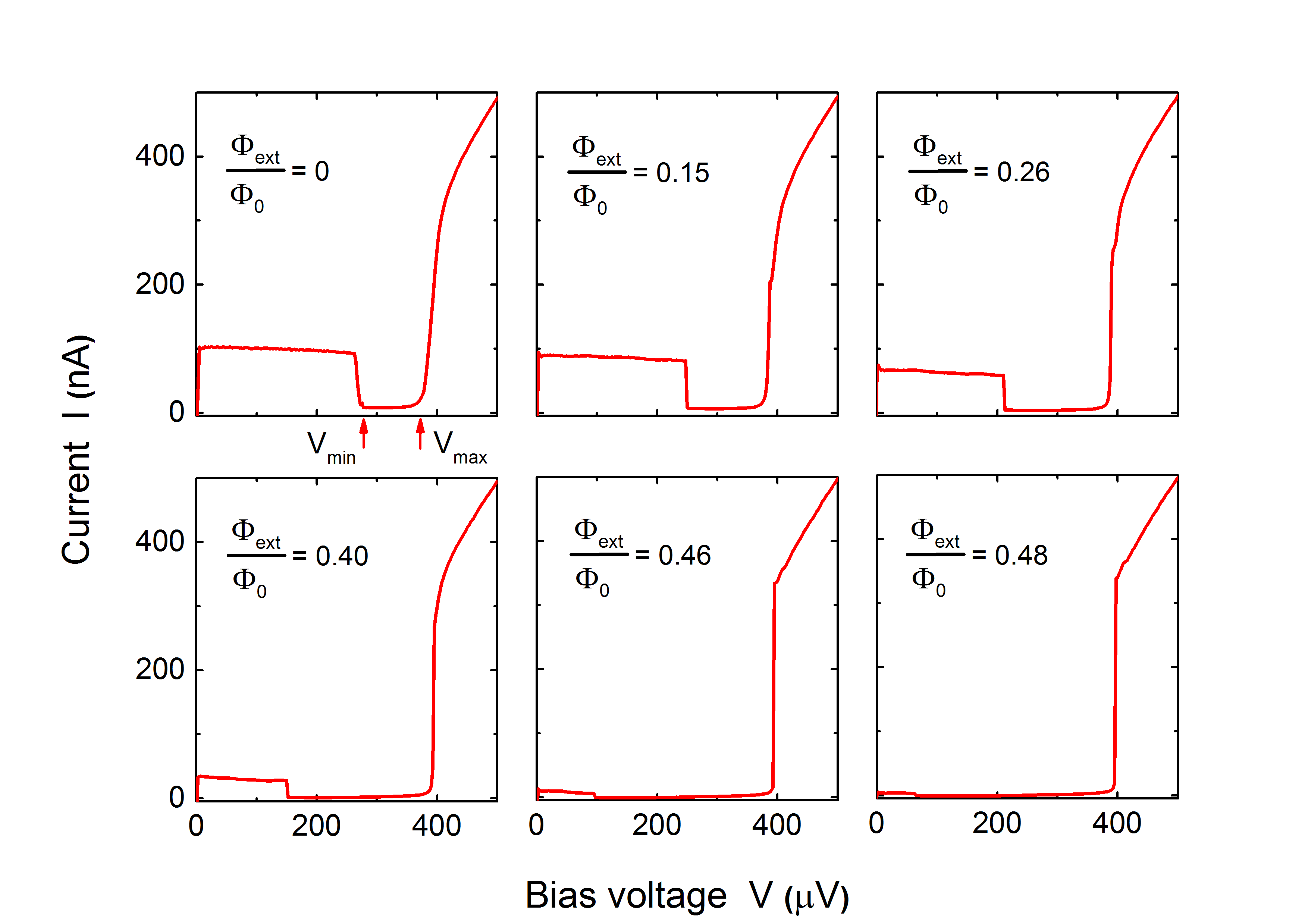}
\caption{(Color online) $I$-$V$ curves measured at $T=\unit[15]{mK}$ in voltage bias mode for device A3 (see table~1 for the parameters) at different values of applied magnetic flux $\Phi_{\rm ext}$.}
\label{fig6IVcurves}
\end{figure}

The tunable range of generation can be extended by varying the magnetic flux penetrating the loop. Figure~\ref{fig6IVcurves} shows a set of $I$-$V$ curves corresponding to different values of magnetic flux $\Phi_{\rm ext}$. One can see that reducing the critical current $I_{\rm C}=I_{\rm C}(\Phi_{\rm ext})$ is accompanied by extending the stable part of the  $I$-$V$ curve down to lower voltages. The amplitude of oscillations, $V_1 \propto I_{\rm C}/\overline{V}$, will be reduced, however, due to the supercurrent suppression.

\begin{figure}[]
\centering%
\includegraphics[width=0.80\columnwidth]{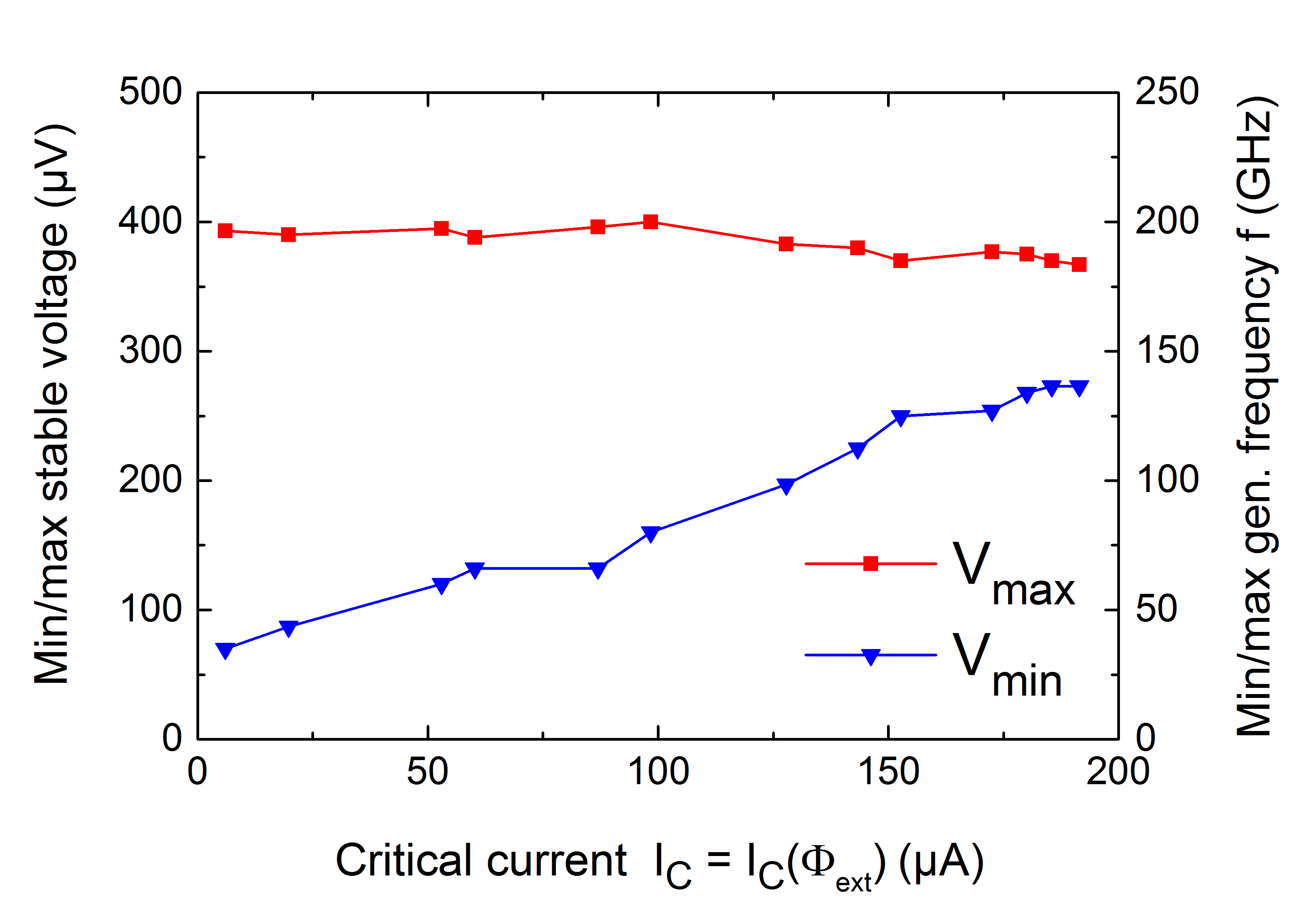}
\caption{(Color online)  Lower ($V_{\rm min}$) and higher ($V_{\rm max}$) voltage limits vs. critical current $I_{\rm C}(\Phi_{\rm ext})$. The data are extracted from the $I$-$V$ curves in figure~\ref{fig6IVcurves}.}
\label{fig7VminVmaxvsB}
\end{figure}

Figure~\ref{fig7VminVmaxvsB} shows a dependence of lower ($V_{\rm min}$) and higher ($V_{\rm max}$) voltage (and frequency) boundaries on the critical current $I_{\rm C}(\Phi_{\rm ext})$, plotted using the data in figure~\ref{fig6IVcurves}. While the value of $V_{\rm max}$ is almost constant around $V_{\rm g}$, $V_{\rm min}$ is decreasing as expected.

\subsection{Results and discussion}

The parameters of the samples as well as their measured properties ($I_{\rm C,max}$=~max[$I_{\rm C}(\Phi_{\rm ext})$], $f_{\rm min,max}=V_{\rm min,max}/\Phi_0 $) are presented in table~\ref{tab1} together with the estimations made for the amplitude and halfwidth of oscillations. In this table, we also estimated the effective junction parameters (\ref{eq:correction Vc beta}--\ref{eq:effective R C}) given by the TJM model and  the electron temperature in thin film resistors $T_{\rm e}$. The values of $V_{\rm C}^\ast$ and $V_{\rm 1}$ were estimated with $I_{\rm C}=I_{\rm C,max}$ and at an average junction voltage $\overline{V} =\unit[300]{\mu V}$. The temperature $T_{\rm e}$  was calculated using (\ref{eq:ElecTemp}) with $\Sigma \sim \unit[10^{10}]{Wm^{-3}K^{-5}}$ \cite{LotkhovUltraHigh}. The value of $T_{\rm e}$ was used for estimating the  halfwidth $\Gamma _{\rm thermal}$ (\ref{eq:linewidth1}) contribution to the total halfwidth $\Gamma _{\rm T}$ (\ref{eq:totalGamma}).

The stable branch of the $I$-$V$ curve and thus the frequency range available for photon generation was found to be limited from above by the steep rise of the quasiparticle current at $V \approx V_{\rm g}$. The extension down to lower voltages (frequencies) was more pronounced for the smaller junctions (type A devices) as compared to the larger ones (type B). On a rough qualitative level, this tendency can be illustrated on the basis of the RSJ model predicting, for $\beta_{\rm C} > 1$, the retrapping current value $I_{\rm min} \propto (1/R^\ast)\sqrt{I_{\rm C}/C}$ (see equation (4.25) in \cite{LikharevBlueBook}) which scales mostly with the oxide barrier transparency and the effective shunt resistance $R^\ast$: the former being in common for both types A and B and the latter lower for type B devices.

\begin{figure}[]
\centering%
\includegraphics[width=1\columnwidth]{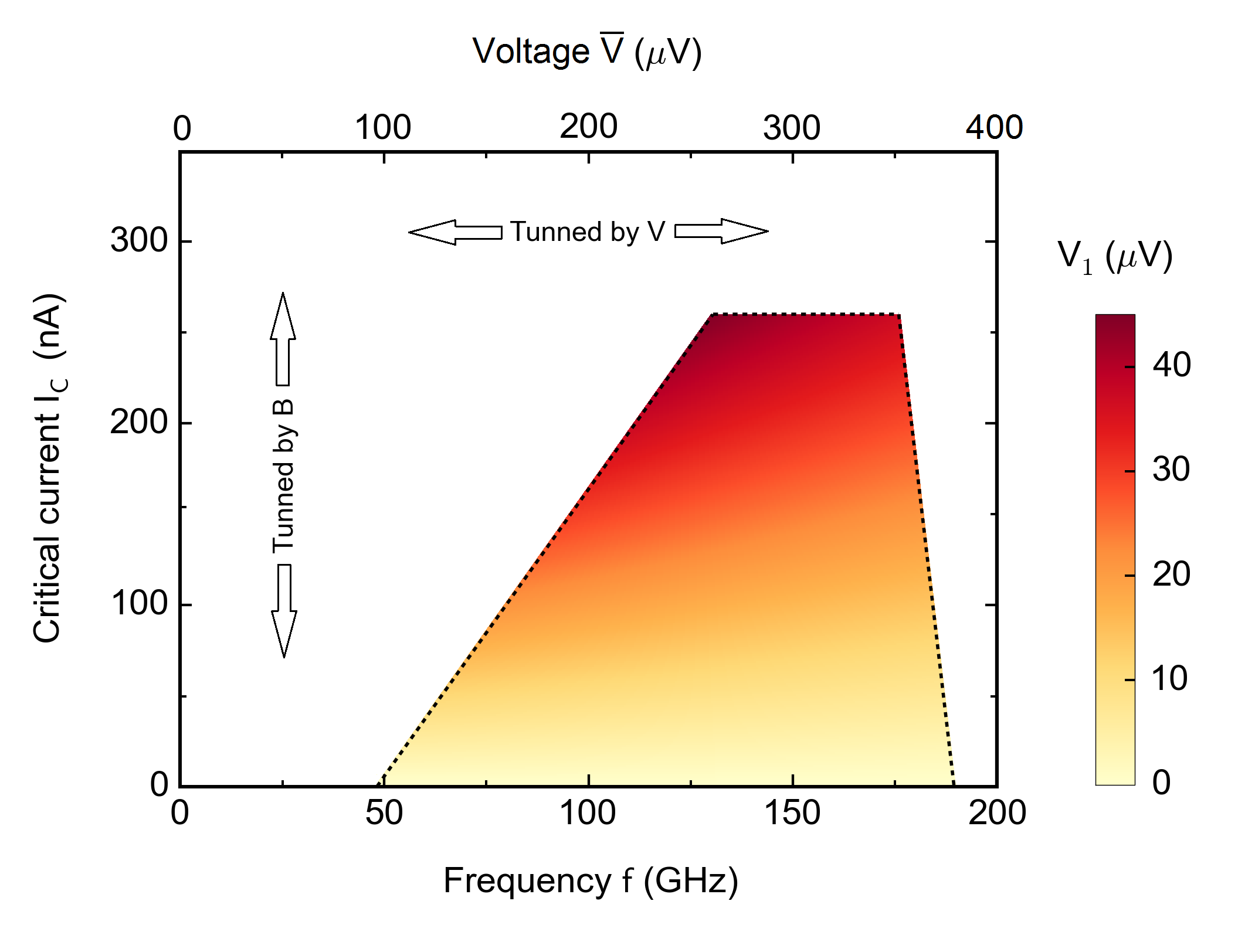}
\caption{(Color online) Estimated amplitude of oscillation $V_1$ as a function of critical current [$I_{\rm C}(\Phi_{\rm ext})$] and frequency $f$ for device A2. Coloured area shows the available region for photon generation.}
\label{fig8FreqRangeandAmp}
\end{figure}

Our evaluation of the amplitude of oscillations resulted in  comparable values for all devices. In the case of $\beta_{\rm C} \gg 1$ and/or $\overline{V} \gg V_{\rm C}$, such a situation becomes evident, owing to the simplified form of  (\ref{eq:amplitude}) reducing to $V_1 \approx I_{\rm C}/(\omega_{\rm J}C)$ and so assuming the amplitude independent of both the junction area and the resistance values. To summarize our results, for one of the junctions (device A2) figure~\ref{fig8FreqRangeandAmp} depicts the estimated amplitude $V_1$ as a function of the critical current and frequency in the range accessible for generation.

The maximum (matched) Josephson radiation power was estimated to be of the order of a few picowatts, which corresponds to the total emission of about $N=P/\hbar\omega_{\rm J}\approx10^{10}$ microwave photons per second (cf., the detection rate of 0.01--10 photons per second mentioned in Sec.~2). Special attention should be paid to designing the microwave coupling element between the generator and the detector in the experiment shown in figure~\ref{fig1GenDet} providing a stable coupling strength over the wide frequency range of interest.

Using thin film resistors of sufficient volume in biasing and measurement leads has provided us with a sufficiently low on-chip heat generation ($T_{\rm e}\sim\unit[150]{mK}$), which has resulted in a narrow linewidth of oscillations of about a few $\unit[]{GHz}$ (which is wider, however, for the larger junctions of type B) and in negligible, almost equilibrium, black-body radiation. The spectrum of black-body radiation is expected  to peak around $\nu_{\rm max} = \unit[0.15]{K} \times W_{\rm k} \approx \unit[9]{GHz} \ll \omega_{\rm J}/2\pi$, where $W_{\rm k} = \unit[58.8]{GHz/K}$ is the Wien's displacement constant,  exponentially decaying into the frequency range of interest.

The evaluation of the total linewidth of oscillations showed a negligibly small contribution of shot noise (\ref{eq:linewidth2}) compared to that of thermal noise (\ref{eq:linewidth1}) for type A devices as a result of a very low quasiparticle current $I_{\rm qp}\approx\unit[10]{nA}$. Thus, for the devices of type A we expect a somewhat smaller linewidth $2\Gamma_{\rm T}$ than for type B devices with a higher value of quasiparticle current $I_{\rm qp}\approx\unit[100]{nA}$, where the contribution of shot noise results in substantial widening of the oscillation linewidth. The widening is more pronounced for circuits with higher resistance values of on-chip resistors (see table 1).

Another important advantage of the small-area junctions appears, again due to considerably lower quasiparticle currents, as a much lower rate of phonon generation $\Gamma_{\rm ph}$. The phonons are mostly generated due to a recombination of quasiparticles created in the leads by current $I_{\rm qp}(V)$ and they can produce a contribution to the detector output \cite{Schinner2009, Gasser2010}, masking the photon signal. Roughly, the rate $\Gamma_{\rm ph}$ can be estimated at its upper limit set by the power dissipated in the junction: $\Gamma_{\rm ph} = \left(\overline{V}\times\overline{I}\right)/2\Delta \sim 10^{10}$ phonons per second. The phonon generation rate appears comparable to that of the photons, but the related energy spectrum is expected \cite{SingerBron1976} to be sharply peaked around the bottom of the excitation band $E_{\rm qp} = 2\Delta \approx h \times \unit[100]{GHz}$, which should help to minimize the phonon signal of the detector at higher frequencies. In any case, a way of mechanical decoupling is currently under development with the purpose of  insulating the microwave detector from acoustic waves produced by the Josephson source. Finally, we note that additional data are also necessary to estimate the effect of substrate piezoelectricity \cite{Koshelets2013}.

\begin{table}

\caption{\label{tab1}Summary of measurement results for $R_{\rm N}$, $R_{\rm S}$, $I_{\rm C,max}$ and $f_{\rm min, max}$ at $T\approx\unit[15]{mK}$ and also estimations for $R^\ast$, ${C^\ast}$, $\beta_{\rm C}^\ast$, $V_{\rm C}^\ast$, $V_{\rm 1}$, $T_{\rm e}$ and $\Gamma _{\rm T}$ for each device.

}

\lineup
\begin{tabular}{@{}lllllllllllll}
\br
No.&$R_{\rm N}$ & $R_{\rm S}$ & $R^\ast$&${C^\ast}^{\rm \dag}$ &$I_{\rm C,max}$ & $\beta_{\rm C}^\ast$ & $V_{\rm C}^\ast$ &$V_{\rm 1}^{\rm\dag}$&$T_{\rm e}^{\rm\dag}$ & $f_{\rm min}^{\rm\ddag}-f_{\rm max}^{\rm\ddag}$&$\Gamma _{\rm T}/2\pi^{\rm\S}$\\\ms
& $(\Omega)$ & $(\Omega)$ &$(\Omega) $&$(\unit{fF})$&$(\unit{\mu A})$ & &$(\unit{\mu V})$&$(\unit{\mu V})$&$(\unit{mK})$&$(\unit{GHz})$&$(\unit{GHz})$\\
\mr
A1 & 690 & \0185 & 154 &\05.3& 0.31 & 0.1 & \048 & 38 & 142 & \060 - 170 & \01.7 \\
A2 & 812 & \0333 & 255 &\05.2& 0.26 & 0.3 & \066 & 42 & 136 & \085 - 190 & \03.0 \\
A3 & 830 & \0975 & 518 &\05.2& 0.19 & 0.8 & \098 & 37 & \091 & \080 - 190 & \06.7 \\
\mr
B1 & \091 & \0186 & \073 &52.3& 2.5 & 2.1 & 184 & 50 & 179 & \095 - 175 & \02.5 \\
B2 & 104 & \0340 & \098 &52.4& 2.6 & 4.0 & 256 & 53 & 166 & 125 - 180 & \04.3 \\
B3 & \087 & \0555 & \096 & 52.0& 2.2 &3.2 & 211 & 45 & 150 & \090 - 170 & \08.2\\
B4 & 107 & 1000 & 125 & 52.4& 2.6 & 6.4 & 325 & 54 & 168 & 125 - 175 & 19.5 \\
\br
\end{tabular}
\noindent $^{\rm \dag}$ Estimated at~$\overline{V}=\unit[300]{\mu V}$\\
\noindent $^{\rm \ddag}$ Measured at~$I_{\rm C}=\unit[0.1]{\mu A}$ or $\unit[1]{\mu A}$\\
\noindent $^{\rm \S}$ Estimated at~$T_{\rm e}$
\end{table}

\section{Conclusion}

We analyzed in detail and realized experimentally a simple Josephson-junction-based generator to be used for future calibration of the single microwave photon detectors based on single electron tunnelling. Corrections to the simplest RSJ model are made within the framework of a more detailed tunnel junction microscopic model. We fabricated a number of Al-based two-junction interferometers embedded in a compact network of on-chip TiO$_{\rm x}$ resistors with a wide range of parameters. DC properties of the $I$-$V$ curves were investigated at $T\approx\unit[15]{mK}$ and the parameters were estimated
in relation to the tunable frequency range, $\omega_{\rm J}/2\pi \sim \unit[60 \div 170]{GHz}$, the emission linewidth and the magnetic field tunable oscillation amplitude. A more detailed characterization of the microwave sources will require a combination experiment for these devices operating together with a single electron tunnelling detector fabricated on the same chip. Further applications, for example in the investigation of superconducting qubits, are expected.

\section*{Acknowledgements}

Technological support from T.~Weimann and V.~Rogalya is appreciated. This work was funded by the Joint  Research  Project  MICROPHOTON. Joint  Research  Project  MICROPHOTON belongs to  the European  Metrology  Research  Programme  (EMRP).  The EMRP is jointly funded by the EMRP participating countries within EURAMET and the European Union.

\end{document}